\documentclass[twocolumn,prl,10pt,nofootinbib,superscriptaddress]{revtex4-1}
\usepackage[utf8]{inputenc}

\def\mySections#1{{\bf #1.} } 
\usepackage{hyperref}
\usepackage{braket}  
\usepackage{upgreek} 
\usepackage[normalem]{ulem}
\usepackage[dvips]{graphicx}
\usepackage{hhline}
\usepackage{epsfig,amsmath,amssymb,verbatim,mathrsfs,array,layout,textcomp,amssymb,latexsym,slashed,graphicx,booktabs,color,mathtools}
\usepackage[caption=false]{subfig} 

\newcommand{\arXiv}[2]{\href{http://arxiv.org/pdf/#1}{{\tt #2/#1}}}
\newcommand{\arXivold}[1]{\href{http://arxiv.org/pdf/#1}{{\tt #1}}}

\newcommand{\beq}{\begin{equation}}
\newcommand{\eeq}{\end{equation}}

\def\beqa{\begin{eqnarray}}
\def\eeqa{\end{eqnarray}}
\def\bea{\begin{eqnarray}}
\def\eea{\end{eqnarray}}

\newcommand{\bv}{\left(\begin{array}{c}}
\newcommand{\ev}{\end{array}\right)}
\newcommand{\bmtwo}{\left(\begin{array}{cc}}
\newcommand{\bmthree}{\left(\begin{array}{ccc}}
\newcommand{\emn}{\end{array}\right)}
\newcommand{\bmtwoc}{\left\{\begin{array}{cc}}
\newcommand{\bmthreec}{\left\{\begin{array}{ccc}}
\newcommand{\emnc}{\end{array}\right\}}
\newcommand{\ba}{\begin{array}}
\newcommand{\ea}{\end{array}}
\newcommand{\be}{\begin{eqnarray}}
\newcommand{\ee}{\end{eqnarray}}

\definecolor{readableRTD}{rgb}{0.7,0.1,0.2}

\def\lsim{\mathrel{\rlap{\lower4pt\hbox{\hskip1pt$\sim$}}
     \raise1pt\hbox{$<$}}}         
\def\gsim{\mathrel{\rlap{\lower4pt\hbox{\hskip1pt$\sim$}}
     \raise1pt\hbox{$>$}}}         

\begin{document}

\title{
Completing the Multi-particle Representations of the Poincar\'e Group
}

\author{Csaba Cs\'aki}
\affiliation{Department of Physics, LEPP, Cornell University, Ithaca, NY 14853, USA}

\author{Sungwoo Hong}
\affiliation{Department of Physics, The University of Chicago, Chicago, IL 60637 , USA }
\affiliation{Argonne National Laboratory, Lemont, IL 60439, USA}
\author{Yuri Shirman}
\affiliation{Department of Physics $\&$ Astronomy, University of California, Irvine, CA 92697, USA }
\author{Ofri Telem}
\affiliation{Theory Group, Lawrence Berkeley National Laboratory, Berkeley, CA 94720, USA}
\affiliation{Berkeley Center for Theoretical Physics, University of California, Berkeley, CA 94720, USA}
\author{John Terning}
\affiliation{QMAP, Department of Physics, University of California, Davis, CA 95616, USA}
\begin{abstract}
We extend the definition of asymptotic multi-particle states of the $S$-matrix, beyond the tensor products of one-particle states. We identify new quantum numbers called pairwise helicities, or $q_{ij}$, associated with asymptotically separated pairs of particles. We first treat all single particles and particle pairs independently, allowing us to generalize the Wigner construction, and ultimately projecting onto the physical states. Our states reduce to tensor product states for vanishing $q_{ij}$, while for vanishing spins they reproduce Zwanziger's scalar dyon states. This construction yields the correct asymptotic states for the scattering of electric and magnetic charges, with pairwise helicity identified as $q_{ij}=e_i g_j -e_j g_i$. 
\end{abstract}
\maketitle


Representations of the Poincar\'e group form the foundation of particle physics, most notably in the construction of the $S$-matrix for quantum-relativistic scattering. The $S$-matrix is the overlap between quantum states representing \textit{free particles} in the asymptotic past and future. This is usually taken to mean that the in- and out- states of the $S$-matrix approach products of one-particle representations of the Poincar\'e group as $t\rightarrow\pm\infty$. Wigner used the method of induced representations to classify the one-particle representations of the Poincar\'e group \cite{Wigner:1939cj} by two properties: their mass and their  little group representation. The little group is the subgroup of Lorentz transformations that leaves a particular reference momentum for a given particle invariant. For massive particles, one can choose the rest frame of the particle, and the little group is the $SU(2)$ double cover of the $SO(3)$ rotations that simply leave the particle at rest. The $2s+1$ dimensional representation of the $SU(2)$ little group is fixed by the spin, $s$, of the particle. For massless particles there is no rest frame, but one can always choose  the reference momentum to point in the $z$-direction. Hence, in this case, the little group is $U(1) \subset ISO(2)$ corresponding to rotations\footnote{As is well known \cite{Wigner:1939cj}, the little group for massless particles is actually larger, forming the 2-dimensional Euclidean group $ISO(2)$. However, picking a nontrivial transformation under the two extra generators in $ISO(2)$ gives rise to continuous single particle quantum numbers, which have not been observed in nature. As is standard, we take the only non-trivially represented generator to be the pure $U(1)$ rotation.} around the $z$-axis. While one-particle representations have been textbook material for many years, surprisingly little attention has been paid to multi-particle representations. In the standard construction of the $S$-matrix, the multi-particle scattering states are simply assumed to be tensor products of one-particle states, i.e. for a two-particle state one simply needs to specify two masses and two spins. Note that multi-particle states are always reducible as representations of the Poincar\'e group, all irreducible representations (irreps) are accounted for by Wigner's method. Nevertheless their decomposition into irreps are not very useful; for example the decomposition of a two-particle state into irreps assigns a different  irrep to each value of the Mandelstam $s=(p_1+p_2)^2$ variable.  
To construct the $S$-matrix one has to properly identify the asymptotic multi-particle states, which are usually assumed\footnote{With the exception of the scattering of particles charged under long-range mediators, we discuss this in more detail later in the paper.} to be tensor products of one-particle states. However, in 1972  Zwanziger pointed out~\cite{Zwanziger:1972sx} that for the scattering of electric and magnetic charges these quantum numbers are not sufficient, and an additional quantum number is needed to characterize the relative transformation of the two-particle state with respect to the simple tensor product state. 

In this paper we present a systematic method to construct the general class of multi-particle states, which automatically include the simple tensor product states, but in addition also contain the generalization of Zwanziger's states. The key insight is the realization that for multi-particle states, in addition to the little groups of each individual particle one also needs to consider the {\it pairwise little groups} corresponding to transformations that leave a pair of momenta invariant \cite{Csaki:2020inw}. Since one can always go into the center of momentum (COM) frame\footnote{For massless particles we can always find a frame where the momenta are collinear.}  of two massive particles, this pairwise little group is also  an $SO(2)\simeq U(1)$ rotation. The corresponding $U(1)$ charge, $q_{ij}$, which we will refer to as pairwise helicity, is required to fully describe the additional phase picked up by a two-particle state under little group  transformations. 
Since there is no Lorentz transformation that leaves three general momenta invariant, the three-particle and higher little groups are all trivial, which implies that the most general multi-particle state is of the form 
\begin{equation}\label{eq:genstate}
|p_1,\ldots , p_n\,;\, \sigma_{i_1}, \ldots , \sigma_{i_n}\,;\,q_{12}, \ldots , q_{n-1,n}\, \rangle
\end{equation}
The aim of this letter is to explain the essence of this construction and establish the main properties of these novel representations. 

The Hilbert space for a single particle is spanned by momentum eigenstates $|p\,;\,\sigma\rangle$ satisfying $P^\mu \,|p\,;\,\sigma\rangle = p^\mu \,|p\,;\,\sigma\rangle$ where $P^\mu$ is the momentum operator and $\sigma$ is a collective index denoting all other quantum numbers besides momentum. To define the precise meaning of these extra quantum numbers, we first have to choose a common reference momentum $k$ for every particle with momentum $p$ and $p^2=m^2$. For massive particles we choose $k=(m,0,0,0)$ while for massless particles we choose $k=(E,0,0,E)$. The \textit{little group} is then defined as the subgroup of Lorentz transformations that leaves $k$ invariant --- which is $U(1)$ (or more precisely $ISO(2)$) for massless particles and $SU(2)$ for massive ones. The label $\sigma$ then serves to fix the transformation of $|k,\sigma\rangle$ under the little group. For example, for massive particles, $\sigma$ stands for both total spin, $s$, and the $z$ component of the spin, $s_z$, and so for all $W\in SU(2)$,
\begin{equation}
U(W)\, |k\,;\,s,\,s_z\,\rangle = D^s_{s'_z s_z} (W)\, |k\,;\, s,\,s'_z\,\rangle~,
\end{equation}
where $D^s_{s'_z s_z}(W)$ is the spin $s$ representation of the $SU(2)$ little group. For massless particles, $\sigma$ stands for helicity $h$, and the little group transformation is just a phase $e^{ih\phi}$, where $\phi$ is the $U(1)$ rotation angle. To obtain the quantum states in a different reference frame, we first define a Lorentz transformation $L_p$ such that $p=L_p k$.  The corresponding quantum state is then defined as $\ket{p\,;\,\sigma}\equiv U(L_p)\ket{k\,;\,\sigma}$. The transformation rule for $\ket{p\,;\,\sigma}$ is then completely fixed by $\sigma$ as follows,
\begin{equation}
U(\Lambda )\, |p\,;\,\sigma \rangle = U(L_{\Lambda p})\,U(W)\, |k\,;\,\sigma \rangle\, ,
\end{equation}
where $W=L_{\Lambda p}^{-1} \Lambda L_p$ takes $k\rightarrow k $ and so is always a little group transformation. Consequently,
\begin{equation}
U(\Lambda )\, |p\,;\,\sigma \rangle = D_{\sigma' \sigma} (W) \,|\Lambda p\,;\,\sigma'\rangle~,
\end{equation}
 where $D_{\sigma' \sigma}$ stands for either $D^s_{s'_z s_z}$ for massive particles or $e^{ih\phi}$ for massless particles. Once the one-particle states are obtained, one can clearly form multi-particle states by considering the tensor product of these states. Surprisingly, these are neither the most general asymptotic multi-particle representations of the Poincar\'e group, nor are they the only ones useful for particle physics. To construct more general $n$-particle representations, we first consider products of $2^n-1$ Poincar\'e groups, and only later focus on their diagonal subgroup, which is the physical Poincar\'e group. Our construction is inspired by the little-group approach to forming on-shell scattering amplitudes, where one initially assigns independent helicity/spin quantum numbers for each spinor-helicity variable, and only at the final step requires that all of these helicities actually describe the transformation under a common (diagonal) Lorentz group. 

For simplicity, we first focus on the construction of two particle states, later generalizing to an arbitrary number of particles. For a pair of particles 1 and 2 we consider representations of the product group $P_1 \times P_2 \times \tilde{P}_{12}$ where each of these $P$'s is a separate copy of the Poincar\'e group. While $P_1$ and $P_2$ correspond to Poincar\'e groups of one-particle states with momenta $p_1$ and $p_2$ respectively, $\tilde P_{12}$ is the group of Poincar\'e transformations acting on a pair of momenta $(\tilde p_1,\tilde p_2)$. At this stage $\tilde p_1$ and $\tilde p_2$ should be thought of as distinct from $p_1$ and $p_2$. Moreover, a momentum pair $(\tilde p_1,\tilde p_2)$ is not a 2-particle state, rather it is simply a tool to define the representations of the Poincar\'e group.  We first define a generalized two particle state as a tensor product of two one-particle state and a  pairwise state 
\begin{equation}
|p_1, p_2, (\tilde{p}_1,\tilde{p}_2)\,;\,\sigma\, \rangle\equiv|p_1\,;\,\sigma_1\rangle\,\otimes\, |p_2\,;\,\sigma_2\rangle\, \otimes\,|(\tilde{p}_1,\tilde{p}_2)\,;\,q_{12}\rangle\,.
\end{equation}
Here $\sigma_i$'s characterize quantum numbers on one-particle states while $q_{12}$ is an extra quantum number associated with the \textit{pairwise helicity}, which we will carefully define below. To obtain representation on physical two-particle states we will eventually restrict $P_1\times P_2\times \tilde P_{12}$ to the diagonal subgroup while projecting onto the physical states via the identification $\tilde p_1=p_1$, $\tilde p_2=p_2$. Note that this projection allows us to interpret  $|(\tilde{p}_1,\tilde{p}_2)\,;\,q_{12}\rangle$  as  a state of two spinless particles with pairwise helicity charge $q_{12}$ ``glued' to spinning particles with momenta $p_1$ and $p_2$.

Similarly to the single particle case, we can again define the relevant reference momenta. For the single particle momenta corresponding to $p_1,\,p_2$, we can choose $k_{1},\,k_2$ defined exactly as in the single particle case. To define the reference momenta $(\tilde{k}_1,\tilde{k}_2)$ corresponding to the pair $(\tilde{p}_1,\tilde{p}_2)$, we go  to the pair's COM frame with the two particles are both moving along the $z$-axis. In this frame we have
\begin{equation}
\tilde{k}_1 = (\tilde{E}_1, 0,0,\tilde{p}_c), \ \ \tilde{k}_2 = (\tilde{E}_2,0,0,-\tilde{p}_c)
\end{equation}
where $\tilde{E}_{1,2}= \sqrt{m_{1,2}^2 +\tilde{p}_c^2}$ and $\tilde{p}_c$ is the Lorentz-invariant COM momentum. The corresponding Lorentz transformations are then
\begin{eqnarray}
&&p_1 = L^1_{p_1} k_1, \quad p_2 = L^2_{p_2} k_2,\nonumber\\[8pt]
&&(\tilde{p}_1~,~\tilde{p}_2) =  \left(\tilde{L}^{12}_{\tilde{p}_1,\,\tilde{p}_2}\,\tilde{k}_1~,~\tilde{L}^{12}_{\tilde{p}_1,\,\tilde{p}_2}\,\tilde{k}_2\right)~.
\end{eqnarray}
Note that unlike the single particle Lorentz transformations, $\tilde{L}^{12}_{\tilde{p}_1,\,\tilde{p}_2}$ takes  $\tilde{k}_1\rightarrow\tilde{p}_1$ \textit{and} $\tilde{k}_2\rightarrow\tilde{p}_2$. This property uniquely determines it, up to a $U(1)$ rotation. 
A generic state is defined by  \vspace*{0.1cm}
\begin{eqnarray}
&& |p_1, p_2, (\tilde{p}_1,\tilde{p}_2)\,;\, \sigma \rangle \equiv \left(\,U( L^1_{p_1} )\,\ket{k_1\,;\,\sigma_1}\right)\,\otimes\nonumber\\
&& \left(U( L^2_{p_2} )\,\ket{k_2\,;\,\sigma_2}\right)\,\otimes\,\left(U( \tilde{L}^{12}_{\tilde{p}_1,\,\tilde{p}_2} )\,\ket{(\tilde{k}_1,\tilde{k}_2)\,;\,q_{12} }\,\right)\, .
 \nonumber
\end{eqnarray}
\\
We can now proceed as Wigner did for the one-particle states by finding the representation of 
Lorentz transformations of the form $\Uplambda\equiv\left(\Lambda_1,\Lambda_2,\tilde{\Lambda}_{12}\right)\in P_1\times P_2\times \tilde{P}_{12}$ on this state \vspace*{0.3cm}
\begin{eqnarray}
&& U(\Uplambda)\,|p_1, p_2, (\tilde{p}_1,\tilde{p}_2), \sigma \rangle = \nonumber\\[10pt]
&&\left(\,D_{\sigma'_1\sigma_1}( W_1)\,\ket{\Lambda_1\, p_1\,;\,\sigma'_1}\right)\,\otimes\left(\,D_{\sigma'_2\sigma_2}( W_2)\,\ket{\Lambda_2\, p_2\,;\,\sigma'_2}\right)\,\otimes\nonumber\\[10pt]&&\left(U(\tilde{L}_{\tilde{\Lambda}_{12}\tilde{p}_1,\,\tilde{\Lambda}_{12}\tilde{p}_2})\,U(\tilde{W}_{12})\,\ket{(\tilde{k}_1,\tilde{k}_2)\,;\,q_{12} }\,\right)\, ,
\end{eqnarray}\\
where 
\begin{eqnarray}\label{eq:Ws}
W_i~&\equiv&~\left(L^{i}_{\Lambda_i p_i}\right)^{-1}\Lambda_i L^i_{p_i}\nonumber\\
\tilde{W}_{12}~&\equiv&~\tilde{L}^{-1}_{\tilde{\Lambda}_{12}\,\tilde{p}_1,\,\tilde{\Lambda}_{12} \tilde{p}_2}\,\tilde{\Lambda}_{12} \,\tilde{L}_{\tilde{p}_1,\,\tilde{p}_2}\, .
\end{eqnarray}
The $W_i$ are just single particle little group transformations, while $\tilde{W}_{12}$ is a transformation that preserves both $\tilde{k}_1$ and $\tilde{k}_2$ defining the  \textit{pairwise} little group. Clearly the pairwise little group is just an $SO(2)\sim U(1)$ rotation around the $z$-axis. Note that the pairwise little group is a true $U(1)$ rotation irrespective of whether the particles are massive or massless. Even in the massless case there is no enhancement of the sort seen for the massless single particle little group, since there is no combination of Lorentz generators that leaves the pair of reference momenta unchanged other than the rotation around the $z$-axis. Defining a rotation angle by $R_z(\tilde{\phi}_{12})\equiv \tilde{W}_{12}$, we then have

\begin{eqnarray}\label{eq:clearrep}
&&U(\Uplambda)\,|p_1, p_2, (\tilde{p}_1,\tilde{p}_2)\,;\, \sigma \rangle=e^{iq_{12}\tilde{\phi}_{12}}\,\cdot\nonumber\\[10pt]
&&D_{\sigma'_1\sigma_1}( W_1)\,D_{\sigma'_2\sigma_2}( W_2)\,|\Lambda_1\,p_1, \Lambda_2\,p_2, (\tilde{\Lambda}_{12}\,\tilde{p}_1,\tilde{\Lambda}_{12}\,\tilde{p}_2)\,;\,\sigma \rangle\, .\nonumber\\
\end{eqnarray}

One can clearly see that Eq.~\ref{eq:clearrep} furnishes a proper representation of the product group $P_1\times P_2 \times \tilde{ P}_{12}$, satisfying all the group product relations by construction. However at this point we still have three separate copies of the Poincar\'e group, and all the momenta $p_1,p_2, {\tilde p}_1,{\tilde p}_2$ are independent. We can now perform a projection onto the physical states, where $p_1={\tilde p}_1, p_2={\tilde p}_2$ and restrict our representation to the diagonal subgroup. 
Now consider the generators of the diagonal subgroup: $P_D^\mu = a \left( P_1^\mu + P_2^\mu + P_{12}^\mu \right)$ and $J_D^{\mu\nu} = b \left( J_1^{\mu\nu} +  J_2^{\mu\nu} +  J_{12}^{\mu\nu} \right)$. One can then easily show that the Lie algebra commutators $[ P_D, P_D ]$ and $[ J_D, J_D ]$ give the correct results for any choices of $a$ and $b$. However, the commutator $[P_D, J_D ]$ requires $b=1$. This freedom allows us to choose $a=\frac{1}{2}$ (corresponding to rescaling positions by a factor of 2) so that the resulting \emph{two-particle} state carries the desired momentum:  
\bea
P_D^\mu \,\vert p_1, p_2, \sigma_1, \sigma_2, q_{12} \rangle = (p_1 + p_2)^\mu\,  \vert p_1, p_2, \sigma_1, \sigma_2, q_{12} \rangle. \nonumber \\
\eea
The transformation of a physical two-particle state is then given by 
\begin{eqnarray}\label{eq:diagonal}
&&U(\Lambda)\,|p_1, p_2\,;\, \sigma_1,\sigma_2\,;\,q_{12} \rangle=\nonumber\\[10pt]
&&e^{iq_{12}\tilde{\phi}_{12}}\,D_{\sigma'_1\sigma_1}( W_1)\,D_{\sigma'_2\sigma_2}( W_2)\,|\Lambda p_1, \Lambda p_2\,;\, \sigma'_1,\sigma'_2\,;\,q_{12} \rangle\, ,\nonumber\\
\end{eqnarray}
with $W_i,\,\tilde{W}_{12}$ given by Eq.~\ref{eq:Ws} with $\Lambda_i=\tilde{\Lambda}_{12}\equiv\Lambda$.

The only remaining task is to check that the action of the unitary operators $U(\Lambda)$ on physical states takes us back to physical states, that is, the projection is preserved under the group transformation. This is obvious from the fact that on the physical states the action of the diagonal Poincar\'e group is $p_1,p_2,(p_1,p_2) \to  \Lambda p_1,\Lambda p_2,(\Lambda p_1,\Lambda p_2)$ hence we clearly remain in the subspace of physical states. Thus we have constructed two-particle representations of the Poincar\'e group, which reduce to the usual tensor product states when $q_{12}=0$. On the other hand, for $q_{12}=1, j_1=j_2=0$ we reproduce Zwanziger's 2-scalar dyon states. Eq.~\ref{eq:diagonal} with $q_{12}$ a half-integer provides the transformation law for the generic 2-particle case. 
The generalization to $n$ particles is now straightforward. We start with a tensor product of $2^n-1$ copies 
\begin{eqnarray}\label{eq:Pmult}
&&P_1\times \ldots \times P_n \times P_{12} \times\ldots \times P_{n-1,n} \times\nonumber\\
&&  P_{123} \times \ldots \times P_{n-2,n-1,n}\times \ldots \times P_{123\ldots n}
\end{eqnarray}
of the Poincar\'e group with each factor $P_{i_1\ldots  i_k}$ represented on an independent $k$-tuple of momenta. However, in 4D, all $k$-tuple little groups are trivial for $k>2$, and so our product group can be represented on states involving only pairwise momenta:
\begin{equation}\label{eq:generalstate1}
|p_1,\ldots, p_n\,;\,(\tilde{p}_1,\tilde{p}_2),\ldots ,(\tilde{p}_{n-2},\tilde{p}_n) ,(\tilde{p}_{n-1},\tilde{p}_n)\,;\,\sigma \rangle\,.~~~~~
\end{equation}
After projecting onto the physical states and reducing to the diagonal subgroup, corresponding to the physical Poincar\'e group, the general transformation rule  becomes

\begin{widetext}
\begin{eqnarray}
U(\Lambda) \, |p_1, \ldots , p_n\,;\,\sigma_1,\ldots ,\sigma_n\,;\,q_{12}, \ldots ,q_{n-1,n} \rangle=\prod_{i>j}e^{i q_{ij} \phi_{ij}}\,
\prod_i\, 
D_{\sigma_i' \sigma_i}(W_i)\,\,
  |\Lambda p_1, \ldots ,\Lambda p_n,\,;\, \sigma_1', \ldots ,\sigma_n'\,;\,q_{12},\ldots q_{n-1,n} \rangle\,. \nonumber \\
\label{eq:diagonalN}
\end{eqnarray}
\end{widetext}

Here is the place to comment on the inequivalence between our newly defined multiparticle representations and standard tensor product representations. To see this inequivalence, note that the projection of the general states Eq.~\ref{eq:generalstate1} to the physical states Eq.~\ref{eq:diagonalN} is injective: different choices of one or two particle representations under the $P_i,\,P_{ij}$ in Eq.~\ref{eq:Pmult} result in different transformation laws in Eq.~\ref{eq:diagonalN}. Consequently, every physical state in Eq.~\ref{eq:diagonalN} can be uniquely lifted to a general one in Eq.~\ref{eq:generalstate1}. The lift is performed by constructing the unique state in Eq.~\ref{eq:generalstate1} which projects to Eq.~\ref{eq:diagonalN}. Clearly, tensor product states are lifted to general states in Eq.~\ref{eq:generalstate1} with all $q_{ij}=0$. By Schur's lemma in the two particle spaces $P_{ij}$, there is no non trivial intertwiner between these states and states with $q_{ij}\neq0$.

Having defined the multi-particle states Eq.~\ref{eq:diagonalN}, we now wish to know if they are a purely mathematical curiosity, or if Nature might have made good use of them. To get a hint for this question, we note that these states possess a unique trait: they involve an extra angular momentum associated with their ``pairwise'' part. This can be read off directly from their definition: the same way that every one of their ``single particle'' components has
\begin{eqnarray}
J_z \, \ket{k_i\,;\,\sigma_i}\,=\,\sigma_{z;i}\, \ket{k_i\,;\,\sigma_i}\, ,
\end{eqnarray}
the ``pairwise'' parts have
\begin{eqnarray}
J_z \, \ket{\tilde{k}_i, \tilde{k}_j\,;\,q_{ij}}\,=\,q_{ij}\, \ket{\tilde{k}_i, \tilde{k}_j\,;\,q_{ij}}\, .
\end{eqnarray}
This extra contribution to the angular momentum is unique in particle physics, as it is associated with particle-pairs, but is independent of the distance between the particles. It is thus a half integer, asymptotic contribution to the total angular momentum. As far as we know, there is only one situation in which this kind of angular momentum is realized: scattering of electric and magnetic charges. Classically, this can be seen by calculating the overall angular momentum stored in the electromagnetic field in the presence of particles with electric (magnetic) charge $e_i$ ($g_i$). The result, in the non-relativistic limit, is given by $\Delta\vec{J} =\sum q_{ij}\hat{r}_{ij}$ where $\hat{r}_{ij}$ is the unit vector between the particles and $q_{ij}=e_i g_j-e_j g_i$ \cite{Schwinger:1969ib,Zwanziger:1968rs}. To construct the relativistic $S$-matrix for electric-magnetic scattering, all we have to do is use the multi-particle states Eq.~\ref{eq:genstate}, identifying pairwise helicity as the half-integer $q_{ij}=e_i g_j-e_j g_i$ \cite{Csaki:2020inw}.

The multiparticle states presented here do not form a Fock space, which by construction is a product space of single particle states. However it has been shown that in 4D  under quite general conditions the scattering of massive particles is described by  multiparticle states in a Fock space\cite{Reed:1979}\footnote{This is to be contrasted to lower dimensions, where anyons and plektons~\cite{Leinaas:1977fm,Wilczek:1982wy} form more general multiparticle states \cite{Mund:1993cf}.}. The way out of this conundrum is to note that the additional asymptotic angular momentum contained in states with non-vanishing pairwise helicity necessarily implies the existence of a massless gauge field under which our asymptotic particles are charged. This violates the underlying assumptions of the theorems requiring a Fock space in 4D, making them non-applicable in our case. Indeed the states introduced in this paper are relevant for describing the scattering of particles interacting with a classical field (which should be described~\cite{Kibble:1968sfb}   
by a so called von-Neumann space \cite{vonNeumann:1938} constructed from an infinite product of Hilbert spaces, since the classical field contains an indefinite number of photons). The traditional approach to describing the $S$-matrix of charged particles sourcing a classical field while avoiding the QED infrared problem \cite{Bloch:1937pw,Kinoshita:1962ur,Lee:1964is,Frohlich:1979uu,Chen:2007ea} is to dress the particles with coherent photon states~\cite{Chung:1965zza,dollard1971}, as done successfully in the Faddeev-Kulish formalism \cite{Kulish:1970ut} (see also \cite{Hannesdottir:2019rqq,Hannesdottir:2019opa} for a recent generalization). Our construction of multiparticle states mirrors the Faddeev-Kulish dressing\footnote{The electric-magnetic Faddeev-Kulish dressing of scattering states was first carried in \cite{Blagojevic:1981he}, where the infrared problem was shown to be solved in a way similar to pure QED. The same formalism was revisited recently in~\cite{Choi:2019sjs}, where the dressing of magnetic sources was shown to involve a 't Hooft line, rather than a Wilson line.}  of multiparticle states for the case when some of the particles are magnetically charged, and so the classical field itself carries asymptotic angular momentum.

Though the formal construction of the electric-magnetic $S$-matrix has been discuss elsewhere \cite{Csaki:2020inw}, we wish to emphasize one unique property. Similarly to the standard construction, the $S$-matrix is defined as 
\beq
S\,=\,\braket{\text{out}|\text{in}},\,
\eeq
where $\ket{\text{in}}\,\left(\,\ket{\text{out}}\,\right)$ are eigenstates of the full Hamiltonian which approach\footnote{These states are actually in the Heisenberg picture, and they approach free states in the sense of time-evolved wave-packets.} free multi-particle states at $t\rightarrow-\infty\,\,\left(t\rightarrow\infty\right)$. While for non-magnetic amplitudes, the corresponding ``free states'' are direct-product multi-particle states, in the electric-magnetic case they are states of the form Eq.~\ref{eq:genstate}. However, there is an added subtlety: as can be seen both classically \cite{Zwanziger:1972sx} and in NRQM \cite{Kazama:1976fm}, the angular momentum of the electromagnetic field flips its sign between the in and out state. To account for that in the electric-magnetic $S$-matrix, the out state has to transform as in Eq.~\ref{eq:diagonalN}, but with $q_{ij}=-(e_i g_j-e_j g_i)$. This is clearly an allowed representation, (we simply change how $q_{ij}$ relates to the charges), and it captures the physics of electric-magnetic scattering \cite{Csaki:2020inw}. It is also the origin of the well-known violation of crossing symmetry for electric-magnetic processes. 

This above result can be easily extended to the case of multiple $U(1)$ gauge groups. There is a well-known generalization \cite{Englert:1976ng,Goddard:1976qe,Weinberg:1979zt} of the `t Hooft-Polyakov monopole construction from $SU(2) \to U(1)$ to $G\to U(1)^n$, where $G$ is a non-Abelian group of rank $\ge n$. The electric and magnetic charges of a particle under each of the unbroken $U(1)$s (corresponding to the Cartan generators of $G$) can be assembled into vectors $\vec{e}$ and $\vec{g}$, both of length $n$. Dirac quantization then requires that $\vec{g}$ is a linear combination of the the simple co-roots of $G$ with integer coefficients \cite{Englert:1976ng}, which guarantees that $\vec{g}\cdot \vec{e}$ is half integer for any $\vec{e}$ descendant from a representation of $G$. For every particle pair, we can now associate pairwise helicity with the Dirac-quantized
\beq
q_{ij}={\vec e}_i \cdot {\vec g}_j-{\vec e}_j \cdot {\vec g}_i~,
\eeq
where the subscript is a particle label.

In this letter we generalized the construction of multi-particle states of the Poincar\'e group beyond the trivial notion of tensor products of $n$ single particle states. In the process, we discovered $n(n-1)/2$ new quantum numbers,  pairwise helicities, denoted by $q_{ij}$. These pairwise helicities provide the pairwise little group phase which determines the Lorentz transformation properties of our multi-particle states with respect to the tensor product states.  Furthermore, we demonstrated how the existence of nonzero pairwise helicity implies an extra contribution to the total angular momentum beyond the orbital and single particle spin/helicity contributions. This new contribution associated with pairs of particles is quantized, and is asymptotically non-decoupling. This leads us to identify our newly defined multi-particle states as the quantum states describing a collection of electric and magnetic charges. Pairwise helicity, in this case, is identified as $q_{ij}=e_i g_j -e_j g_i$, which is half-integer by virtue of Dirac-Zwanziger-Schwinger quantization.  
 
\quad\\
 
\mySections{Acknowledgments}
The authors would like to thank Michael Waterbury for helpful discussions. We would also like to thank the referee for pointing out essential literature concerning multiparticle scattering.
 C.C.  was supported in part by the NSF grant PHY-2014071 as well as the US-Israeli BSF grant 2016153.  
 S.H. is supported by a DOE grant DE-SC-0013642 and a DOE grant DE-AC02-06CH11357. 
 Y.S. is supported in part by the NSF grant  PHY-1915005.
J.T. is supported by the DOE under grant  DE-SC-0009999. OT is supported in part by the DOE under grant DE-AC02-05CH11231.

\bibliographystyle{apsrev4-1}

\end{document}